\documentclass[aps,prl,twocolumn,superscriptaddress,showkeys,nofootinbib,notitlepage,longbibliography]{revtex4-1}
\usepackage[english]{babel}
\usepackage{xcolor}
\usepackage{mathtools}
\usepackage{pdfsync}
\usepackage{amssymb}
\usepackage{amsmath}

\usepackage{slashed}
\usepackage[active]{srcltx}

\def\d{\hbox{{d}\kern-.20em\hbox{l}}}

\def \matrix #1 {\left(\begin{array}{cc} #1 \end{array}\right)}

\def\II{\hbox{{1}\kern-.25em\hbox{l}}}

\newcommand{\cusp}{{\scriptscriptstyle \rm cusp}}
\newcommand{\MS}{\overline{\scriptscriptstyle \rm MS}}

\begin{document}


\title{
\hfill {\textmd{DESY 19--080}}
\\[3mm]
Two-loop evolution equation for the $B$-meson distribution amplitude}


\author{V. M. Braun}
   \affiliation{Institut f\"ur Theoretische Physik, Universit\"at
   Regensburg, D-93040 Regensburg, Germany}

   \author{Yao Ji}
\affiliation{Institut f\"ur Theoretische Physik, Universit\"at
   Regensburg, D-93040 Regensburg, Germany}

  \author{A. N. Manashov}
\affiliation{ Institut f\"ur Theoretische Physik, Universit\"at Hamburg
   D-22761 Hamburg, Germany}
   \affiliation{Institut f\"ur Theoretische Physik, Universit\"at
   Regensburg, D-93040 Regensburg, Germany}
\affiliation{
   St.Petersburg Department of Steklov
Mathematical Institute, 191023 St.Petersburg, Russia}


\begin{abstract}
We derive the two-loop evolution equation of the B-meson light-cone distribution amplitude
which is the last missing element for the next-to-next-to-leading logarithmic resummation of
QCD corrections to B decays in QCD factorization.
We argue that the evolution kernel to all orders in perturbation theory can be written as a
logarithm of the generator of special conformal transformations
times the cusp anomalous dimension, up to a scheme-dependent overall constant.
Up to this constant term,
the evolution kernel to a given order in perturbation theory can be obtained from the calculation
of special conformal anomaly at one order less.
\end{abstract}


\keywords{B decays, QCD factorization, Conformal symmetry}

\maketitle

%
%

The B-meson light-cone distribution amplitude (LCDA)~\cite{Grozin:1996pq}
is the crucial nonperturbative quantity in the description of charmless hadronic B-decays
and studies of direct CP violation in the framework of QCD
factorization~\cite{Beneke:1999br,Beneke:2000ry,Beneke:2001ev} and the  ``perturbative QCD'' (pQCD)
factorization~\cite{Keum:2000wi,Lu:2000em,Li:2009wba}.
It is also the central element in B-decay form factor calculations using various techniques.
In particular the leptonic radiative decay $B\to\ell\nu_\ell\gamma$
is generally viewed as the theoretically cleanest process from where the information
on the B-meson LCDA can eventually be extracted with the least uncertainties, see
\cite{Beneke:2011nf,Beneke:2018wjp,Shen:2018abs} for the recent developments.
The related studies constitute a large fraction of the Belle II physics program~\cite{Kou:2018nap}.
Having in mind very high statistical accuracy of the expected data it is imperative
to make theory description as precise as possible.

As it is common in field theories, extraction of the asymptotic behavior --- here the heavy quark limit ---
produces divergences that have to be renormalized,
so that the B-meson LCDA is scale- and scheme-dependent.
The corresponding one-loop evolution equation was derived in Ref.~\cite{Lange:2003ff}.
This equation has an interesting structure related to the symmetry of the problem under
special conformal transformations (inversion with respect to the heavy quark position,
infinitesimal translation along its four-velocity vector and the second inversion~\cite{Knodlseder:2011gc}).
This symmetry allows one to obtain the analytic expression~\cite{Bell:2013tfa,Braun:2014owa} for the
eigenfunctions and the anomalous dimensions.

In this work we argue that the structure found in Ref.~\cite{Braun:2014owa} holds to all orders
in perturbation theory: The evolution kernel $\mathcal{H}(a)$, $a = \alpha_s/(4\pi)$,
(precise definition will be given below) can be written as a logarithm of
the generator of special conformal transformation $\mathcal{K}(a)$ times the cusp
anomalous dimension $\Gamma_{\rm cusp}(a)$, up to an overall additive
constant
\begin{align}
 \mathcal{H}(a) &= \Gamma_{\cusp} (a) \ln (i \mathcal{K}(a) \tilde \mu e^{\gamma_E}) +
\Gamma_{\!\scriptscriptstyle +}(a)\,.
\label{all-order}
\end{align}
Here and below $\tilde\mu = \mu^{\scriptscriptstyle \overline {\rm MS}}\,e^{\gamma_E}$. Apart from an elegant interpretation of the
solutions --- eigenfunctions of the B-meson LCDA evolution equation are eigenmodes of special conformal transformations ---
utility of this representation is that the nontrivial part of the evolution equation at any given order in perturbation theory can
be obtained by the calculation of special conformal anomaly (quantum deformation of $\mathcal{K}$) at one order less. We have
verified this result by explicit calculation to the two-loop accuracy. The resulting two-loop evolution equation
\eqref{s-space-evol}
is directly
relevant for phenomenology and allows one, e.g., to perform a complete
next-to-next-to-leading-logarithmic (NNLL)
resummation of heavy quark mass logarithms in the
$B\to\ell\nu_\ell\gamma$  decay.

We start with a summary of the one-loop results.
The $B$-meson LCDA is defined~\cite{Grozin:1996pq} as a matrix element of
the renormalized light-ray operator
\begin{align}\label{Oz}
\mathcal{O}(z)=\bar q(zn) \slashed{n}\gamma_5 h_v(0),
\end{align}
built of a heavy quark field  $h_v(0)$ in effective theory (HQET) and a light anti-quark $\bar q(zn)$,
between the vacuum and $B$-meson state
\begin{align}
\langle 0| \mathcal{O}(z)|\bar B(v)\rangle &=
i F(\mu) \Phi_+(z,\mu)
\notag\\&=
i F(\mu) \int_0^\infty\!d\omega\, e^{-i\omega z} \phi_+(\omega,\mu)\,.
 \end{align}
Here $v_\mu$ is the heavy quark velocity,
$n_\mu$ is a light-like vector, $n^2=0$, and we assume that $n\cdot v=1$.
The Wilson line connecting the fields is tacitly implied.
The operator in Eq.\,\eqref{Oz} is assumed to be renormalized in the
$\overline{\text{MS}}$ scheme, $\mu = \mu^{\scriptscriptstyle \overline {\rm MS}}$ is the factorization scale
and $F(\mu)$ is the HQET B-meson decay constant.
The corresponding anomalous dimension is~\cite{Neubert:1993mb}
\begin{align}\label{eq:gammaF}
 \gamma_F(a) &= -3 a C_F + a^2 C_F\biggl\{C_F\Big[\frac52-\frac{8\pi^2}{3}\Big]
\notag\\&\hspace*{2.7cm} + C_A \Big[1+ \frac{2\pi^2}{3}\Big] -\frac52 \beta_0\biggr\}.
\end{align}

For the most part of this work it will be convenient to stay in position space.
The scale dependence of  $\Phi_+(z,\mu)$ is governed by the renormalization group (RG) equation
for the nonlocal operator $\mathcal O(z)$ which has the form
\begin{align}\label{RGE}
 \Big(\mu\frac{\partial}{\partial\mu}+\beta(a) \frac{\partial}{\partial a}+\mathcal{H}(a)\Big)
\mathcal O(z)=0\,,
\end{align}
where $\beta(a)$ is the QCD-beta function,
$a=\alpha_s/4\pi$ and $\mathcal{H}(a)=a\mathcal{H}^{(1)}+a^2 \mathcal H^{(2)}+\ldots$
is an integral operator (evolution kernel).
The leading term $\mathcal H^{(1)}$ was calculated by Lange and
Neubert~\cite{Lange:2003ff}. Their result converted to position space takes the form~\cite{Braun:2003wx,Knodlseder:2011gc}
\begin{align}\label{Honeloop}
\mathcal H^{(1)} \mathcal O(z) &=4C_F\Big\{ \big[\ln(i \tilde\mu z)-1/4\big]\mathcal O(z)
\notag\\
&\quad
+\int_0^1 du \frac{\bar u }u \big[\mathcal{O}(z)-\mathcal O(\bar u  z)\big]\Big\},
\end{align}
where $\bar u  = 1-u $.

It turns out that this expression (apart from the constant term -1/4) can be found without calculation and
is fixed by the symmetry of the problem. We remind that QCD Lagrangian is conformally invariant
at the classical level, and as a consequence one-loop evolution kernels for composite operators built
from light quarks commute with the generators of conformal transformations. It is, therefore, possible
to write these kernels as \emph{functions} of the quadratic Casimir operator of the
collinear subgroup~\cite{Braun:2003wx}.
For the heavy-light operators considered here the conformal symmetry is lost because the effective
heavy-quark field $h_v$ is essentially a nonlocal object --- it can be replaced by the
Wilson line going from zero to infinity along the velocity vector $v^\mu$~\cite{Korchemsky:1991zp}
--- and it does not transform covariantly under the Poincare group.
A special conformal transformation in the direction  $v^\mu$
is an exception as it leaves the $v$-ordered Wilson line (and the light-like one) invariant. Thus one should expect that
\begin{align}\label{K}
[K,\mathcal H^{(1)}]=0\,,
\end{align}
where $K=v^\mu K_\mu$, and $K_\mu$ is the generator of special conformal transformations.
 The dilatation invariance of the evolution kernel is also lost because of the term $\sim \ln i\mu z$ that is due to the cusp
in the Wilson line between the light-like (in the direction of $n^\mu$) and time-like (in the direction of $v^\mu$) segments. The
coefficient in front of $\ln i\mu z$  is called cusp anomalous dimension~\cite{Korchemsky:1987wg} and is known at NNLO
\cite{Moch:2004pa},
\begin{align}
  \Gamma_\cusp(a) &= a \Gamma_\cusp^{(1)} + a^2 \Gamma_\cusp^{(2)} +\ldots
\\&=
4 C_F a + \frac43 C_F a^2 \Big[(4\!-\!\pi^2)C_A + 5 \beta_0\Big] +\ldots\notag \, .
\end{align}
To one-loop accuracy one obtains therefore
\begin{align}\label{D}
[D,\mathcal H^{(1)}] &=  \Gamma_\cusp^{(1)}  = 4C_F\,.
\end{align}
Eq.\,\eqref{K} implies that the operators ${\cal H}^{(1)}$ and $K$ can be diagonalized simultaneously, Since the problem has one
degree of freedom, this means that the evolution kernel can be written as a \textit{function} of $K$,
${\cal H}^{(1)} = f(K)$. This function is fixed by Eq.\,\eqref{D}
and the canonical commutation relation $[D,K] = K$ which implies that, for arbitrary power $m$,  $[D,K^m] = m K^m$.
Thus
\begin{align}
  [D, f(K)] = K \frac{\partial}{\partial K} f(K) =  \Gamma_\cusp^{(1)},
\end{align}
so that  $f(x) = \Gamma_\cusp^{(1)} \ln x +\, \text{const}$. The integration constant remains
undetermined and has to be calculated explicitly.  One obtains~\cite{Braun:2014owa}
\begin{align}
\label{H1}
  \mathcal{H}^{(1)} & =  \Gamma_\cusp^{(1)} \ln\big( i\tilde\mu e^{\gamma_E} K\big) - 5 C_F\,.
\end{align}
Note that the derivation only uses the commutation relations for the generators.

The dilatation and conformal symmetry generators in position space are simple first-order
differential operators
\begin{align}\label{DKvariation}
D \,\mathcal O(z)=(z\partial_z+3/2)\,\mathcal O(z)\,,
\notag\\
K \mathcal O(z)=(z^2\partial_z+2z)\,\mathcal O(z)\,,
\end{align}
which coincide (up to the replacement $1\mapsto 3/2$ in $D$) with the generators  $S_0$ and $S_+$
of the collinear subgroup, respectively~\cite{Braun:2003rp,Knodlseder:2011gc}.
Using these expressions one can verify~\cite{Braun:2014owa} that the representation in Eq.\,\eqref{H1}
is indeed equivalent to Eq.\,\eqref{Honeloop} obtained by explicit calculation.
Moreover, eigenfunctions of $K$ are easy to find:
\begin{align}
\label{Qs}
 Q_s(z) = -\frac{1}{z^2} e^{is/z}, \quad iK\,Q_s(z) = s Q_s(z)\,,
\end{align}
where $s\ge 0$ to ensure analyticity in the lower half-plane~\cite{Braun:2003wx}.
They provide the basis of the eigenfunctions for
the (one-loop) evolution kernel
\begin{align}
 \mathcal{H}^{(1)}  Q_s = \Big[\Gamma_\cusp^{(1)} \ln\big( \tilde\mu e^{\gamma_E} s \big) - 5 C_F\Big] Q_s\,.
\end{align}
Thus one can write the LCDA as an integral~\cite{Braun:2014owa}
\begin{align}
\label{s-representation}
  \Phi_+(z,\mu) = \int_0^\infty \!ds\, s\,\eta_+(s,\mu)\,Q_s(z)\,,
\end{align}
where functions $\eta_+(s,\mu)$ are multiplicatively renormalizable. The corresponding momentum-space
expression is in terms of Bessel functions~\cite{Braun:2014owa}. The representation \eqref{s-representation}
is equivalent to the one
found in Ref.~\cite{Bell:2013tfa}.

In this work we argue that the similar representation of the evolution kernel, Eq.\,\eqref{all-order},  holds to
all orders in perturbation theory, where all three elements: $\Gamma_\cusp(a)$,
$\Gamma_+(a)$, and the generator of special conformal transformations $\mathcal{K}(a)$
include higher-order corrections.

The starting observation is that the RG kernels in the $\overline{\text{MS}}$ scheme do not depend
on $\epsilon =(4-d)/2$ by construction. They are, therefore, the same for QCD in $d=4$ and
in $d=4-2\epsilon$ dimensions at the critical point $a=a_*$ where $\beta(a_*)=0$ and the theory
enjoys exact scale and conformal invariance~\cite{Braun:2018mxm}. This ``hidden symmetry'' of QCD evolution
equations was identified and applied before to the study of the leading twist-operators
to three-loop accuracy~\cite{Braun:2016qlg,Braun:2017cih}.

Generators of symmetry transformations acting on composite operators in an interacting theory
are, generally, modified by quantum corrections~\cite{Braun:2014vba,Braun:2016qlg,Braun:2019qtp}
\begin{align}\label{calDK}
\mathcal{D}(a_*)&= D -\epsilon +\mathcal H(a_*)\,,
\notag\\
\mathcal{K} (a_*)&= K  - \epsilon z +  z \Delta(a_\ast) \,,
\end{align}
where $D,K$ are the corresponding canonical expressions \eqref{DKvariation} and
$\epsilon=\epsilon(a_*)=-\beta_0 a_*+O(a_*^2)$.
Note that the generator of dilatations $\mathcal{D}(a_*)$ can be written in terms of
the RG kernel, whereas the generator of special conformal transformations $\mathcal{K} (a_*)$
cannot be fixed from general considerations and contains a correction term $\Delta$. It can be calculated in perturbation
theory $\Delta(a_\ast) = a_\ast \Delta^{(1)} + a_\ast^2 \Delta^{(2)} + \ldots$ using conformal Ward identities, see
Refs.~\cite{Mueller:1991gd,Belitsky:1998gc,Braun:2016qlg} for a detailed discussion.

The modified generators obey, by definition, the same canonical commutation relation
\begin{align}\label{eq:DKcomm}
[\mathcal{D}(a_*),\mathcal{K}(a_*)] = \mathcal{K}(a_*)
\end{align}
whereas
Eqs.~\eqref{K},\eqref{D} are generalized to
\begin{subequations}\label{algebra}
\begin{align}
 \label{algebraK}
 [\mathcal{K}(a_*), \mathcal{H}(a_*)] & =0\,,
\\
 \label{algebraD}
 [\mathcal{D}(a_*), \mathcal{H}(a_*)] & =  [D, \mathcal{H}(a_*)] = \Gamma_{\cusp}(a_*)\,.
\end{align}
\end{subequations}
The second relation follows from the known result~\cite{Korchemsky:1987wg} that the {$\ln (i\mu z)$} term can appear in $\mathcal{H}(a_*)$ only linearly
(to all orders in perturbation theory) and its coefficient is the cusp anomalous dimension.
Note that in contrast to $\mathcal H$ the correction term $\Delta$ in the generator of special conformal transformation
does not contain $\sim \ln \mu z$ contributions:  Using Eqs.\,\eqref{calDK}--\eqref{algebra}
one can show that $[z\partial_z,\Delta]=0$. This means that $\Delta$ can be written as a function of $(z\partial_z)$ and
rules out possibility of logarithmic contributions.
The representation for $\mathcal{H}$ in Eq.\,\eqref{all-order} follows from the commutation relations \eqref{algebra}
in the same way as the one-loop expression \eqref{H1} follows from \eqref{K} and \eqref{D}.

Aiming at the two-loop accuracy for the evolution kernel one needs, obviously, a one-loop correction to $\mathcal{K}$.
A straightforward calculation (cf. \cite{Braun:2016qlg}) gives
\begin{align}\label{Delta-explicit}
\Delta^{(1)}\mathcal O(z)& = C_F\biggl\{3 \mathcal O(z) + 2\! \int_0^1\!\! du \,w(u)
\big[ \mathcal O(z)-\mathcal O(\bar u z)\big] \biggr\},
\notag\\
w(u)&={2\bar u}/u+\ln u\,.
\end{align}
We have checked that the same result can be obtained starting from the one-loop correction to the generator of special conformal
transformations for the light-quark system~\cite{Belitsky:1998gc,Braun:2014vba,Braun:2016qlg} and applying the
``light-to-heavy'' reduction procedure suggested in Ref.~\cite{Braun:2018fiz}. Thus in fact a new calculation is not needed.

For practical applications,  explicit expression for the kernel as an integral operator,
similar to the one-loop result in Eq.\,\eqref{Honeloop}, can be more useful.
To find this expression one can use the following ansatz
\begin{align}
 \mathcal{H}^{(2)} & = \Gamma^{(2)}_{\cusp}\, H_1  + \Gamma^{(1)}_{\cusp}\, \delta H +\,\text{const}\,,
\notag\\
   \delta H\, \mathcal{O}(z) &= \int_0^1 du \,{\frac{\bar u}u}\, h(u)\big[ \mathcal O(z)-\mathcal O(\bar u z)\big]\,,
\end{align}
where $H_1$ is the one-loop kernel \eqref{Honeloop} stripped of the $4C_F$ factor, so that $[D,H_1]=1$, $[K,H_1]=0$, $[D,\delta H]=0$.
In this way Eq.\,\eqref{algebraD} is fulfilled identically
and the function $h(u)$ can be found from Eq.\,\eqref{algebraK}.
To this end it is convenient to write $\mathcal{K} = z(D+1/2-\epsilon + \Delta)$
and to the required accuracy replace $\epsilon\mapsto -\beta_0 a_\ast$.
Working out the commutators and using that
\begin{align}
[\Delta^{(1)}, H_1] & =  [\Delta^{(1)},\ln z], &&  [z,H_1]=-z\frac{1}{z\partial_z+2},
%
\end{align}
one obtains after some algebra
\begin{align}\label{aaa}
[\delta H, z] &= z \mathbb{T}  \Big\{ [\Delta^{(1)},\ln z] - \mathbb{T} \big(\beta_0+\Delta^{(1)}\big) \Big\},
\end{align}
where the operator $\mathbb{T}$ is defined as
\begin{align}
\mathbb{T}\,  \mathcal{O}(z)& = \frac{1}{z\partial_z+2}\, \mathcal{O}(z)
= \int_0^1du\,\bar u\, \mathcal O(\bar u z)\,.
\end{align}
The remaining commutators are:
\begin{align}\label{primes}
 [\Delta^{(1)},\ln z] \mathcal{O}(z)&=- 2 C_F \int_0^1du \,\ln(\bar u)\, w(u)\, \mathcal{O} (\bar u z),
\notag\\
[\delta H, z] \mathcal{O}(z)  &= z \int_0^1du\,\bar u\, h(u)\, \mathcal{O} (\bar u z).
\end{align}
Using these expressions and \eqref{Delta-explicit} in \eqref{aaa}, we obtain
\begin{align}
\label{result2}
    h(u) &=  \ln\bar u \Big[\beta_0 +
 2 C_F \Big(\ln\bar u -\frac{1\!+\!\bar u}{\bar u}\ln u-\frac32\Big)  \Big].
\end{align}
Collecting all terms one gets
\begin{align}
\label{result1}
  \mathcal{H}(a)\mathcal{O}(z) &= \Gamma_{\cusp} (a)\biggl\{
\ln (i\tilde\mu z) \mathcal{O}(z) +  \int_0^1\!\!du \,\frac{\bar u}{u} [1+a h(u)]
\notag\\&\quad
\times \big[\mathcal{O}(z) - \mathcal{O}(\bar u z)\big] \biggr\} + \gamma_{+}(a)\,.
\end{align}
The constant $\gamma_{+}(a)$ requires explicit calculation (see below).
We obtain
\begin{align}
\label{result3}
 \gamma_{+}(a) &= -a C_F +  a^2 C_F
\biggl\{
4 C_F \left[\frac{21}{8} + \frac{\pi^2}{3} - 6\zeta_3\right]
\notag\\&
+ C_A \left[\frac{83}{9} -\frac{2\pi^2}{3} - 6\zeta_3\right]
+ \beta_0\left[\frac{35}{18} -\frac{\pi^2}{6}\right]
\biggr\}.
\end{align}
The anomalous dimension $\Gamma_+(a)$ appearing in Eq.\,\eqref{all-order} is given by
\begin{align}\label{eq:Gamma+}
 \Gamma_+(a) &= \gamma_+(a) - \Gamma_{\cusp}(a) \big[1-a\, \varkappa + \mathcal{O}(a^2)\big]\,,
\notag\\
\varkappa  &= \int_0^1\!\! du \frac{\bar u}{u} h(u) = C_F\Big[\frac{\pi^2}{6}\!-\!3\Big]
+ \beta_0 \Big[1\!-\!\frac{\pi^2}{6}\Big].
\end{align}

The result can also be cast in the form of an equation for the scale dependence of the coefficients in the expansion
\eqref{s-representation} of the
LCDA in the eigenfunctions \eqref{Qs} of the one-loop evolution equation
\begin{align}
& \Big(\mu\frac{\partial}{\partial\mu}+\beta(a) \frac{\partial}{\partial a}
  + \Gamma_{\cusp}(a) \ln \big(\tilde \mu e^{\gamma_E} s \big) + \gamma_\eta(a)\Big) \eta_+(s,\mu)
\nonumber\\
 & \hspace*{2.5cm}{}= 4 C_F a^2 \int_0^1\!du \frac{\bar u }{u}h(u) \eta_+(\bar u s,\mu)\,,
\label{s-space-evol}
\end{align}
where the kernel $h(u)$ is given in Eq.\,\eqref{result2} and
$\gamma_\eta(a) = \Gamma_+(a) - \gamma_F(a)$, (see Eq.~\eqref{eq:gammaF}, \eqref{result3}, and \eqref{eq:Gamma+}).

In order to derive the expression for $\gamma_+(a)$ in Eq.\,\eqref{result3}, and also for independent verification of Eq.\,\eqref{result1}
obtained from symmetry considerations, we have calculated the two-loop kernel ${\mathcal H}^{(2)}$ explicitly.
The contributing Feynman diagrams can be split into three classes: ``light vertex'', describing the interaction of the light
antiquark with the light-like Wilson line, ``heavy vertex'', the same but for the heavy quark, and ``exchange'' diagrams, involving
interaction between the heavy quark and the light antiquark. The answers for the two-loop light vertex diagrams can be found
in Appendix C of Ref.~\cite{Braun:2016qlg}. The sum of heavy vertex diagrams has the form
$\Gamma^{(2)}_\cusp\ln(i\tilde\mu z) +\, \text{const}$, and the constant term is the one of interest. The calculation of exchange
diagrams is considerably simplified thanks to the one-loop exchange diagram being finite~\cite{Grozin:1996pq}.
It turns out that the two-loop heavy-light exchange diagrams can be obtained from the
expressions for their light-light counterparts collected in Appendix C of Ref.~\cite{Braun:2016qlg}
by throwing out all terms where the heavy quark is moved from the origin in position space.
The results for separate diagrams will be presented elsewhere.

%
\begin{figure}[t]
\centering
 \includegraphics[width=0.45\textwidth]{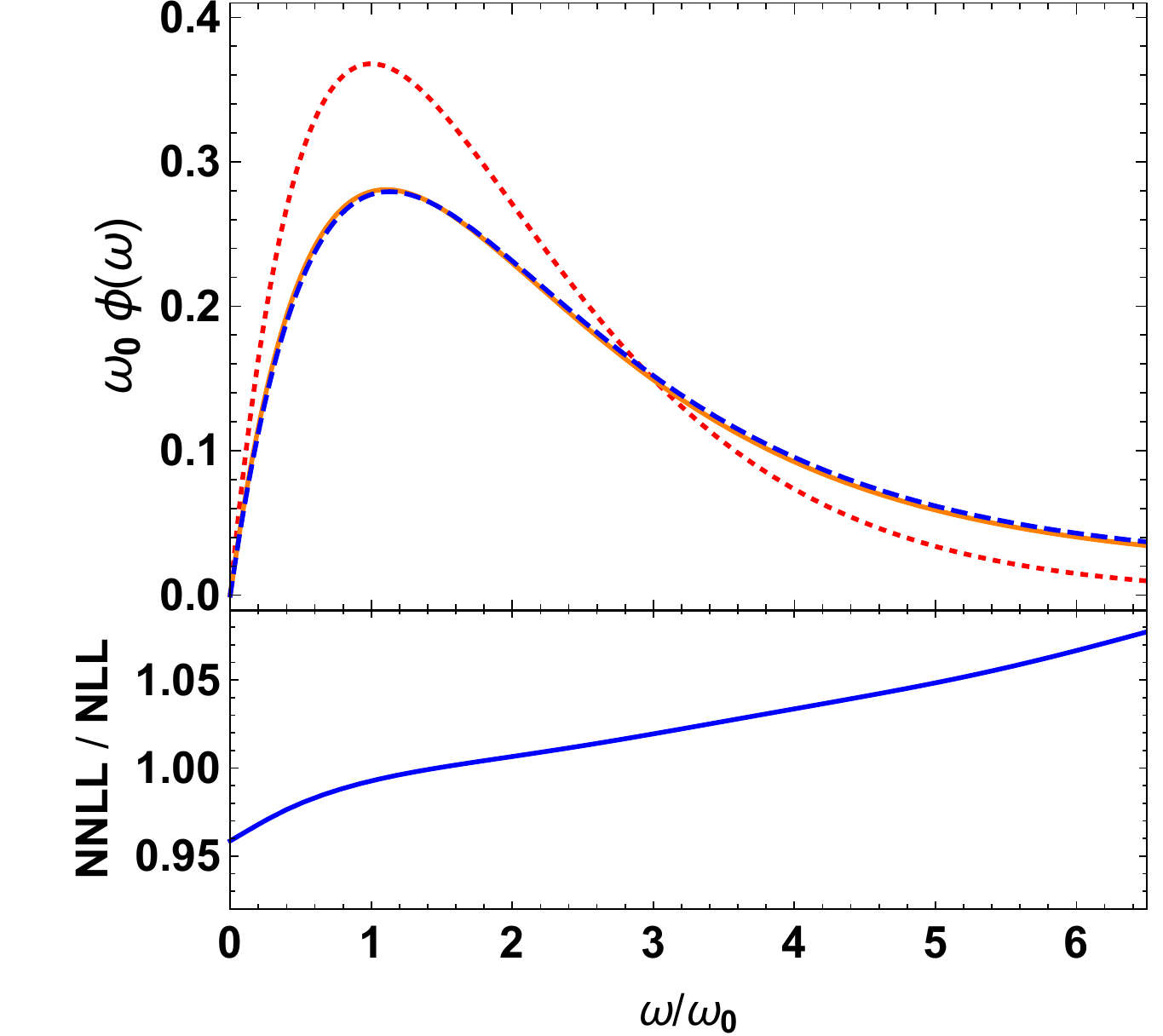}
\caption{The B-meson LCDA \eqref{GNmodel} at the reference scale $\mu_0^{\MS}=1\,\text{GeV}$ (red dots)
and after the evolution to $\mu^{\MS}=2\,\text{GeV}$ with the NLL (blue dashes) and
NNLL (red solid) accuracy.
}
\label{fig:GNmodel}
\end{figure}

The size of the two-loop correction is illustrated in Fig.~\ref{fig:GNmodel}
for the simplest one-parameter exponential model of the LCDA at the reference
scale $\mu_0^{\MS}=1\,\text{GeV}$~\cite{Grozin:1996pq}
\begin{align}
 \phi_+(\omega,\mu_0) &= \frac{\omega}{\omega_0^2} e^{-\omega/\omega_0}.
\label{GNmodel}
\end{align}
For this plot we take $\omega_0 = 300\,\text{MeV}$.
We show the LCDA at the reference scale and after evolution to
$\mu^{\MS}=2\,\text{GeV}$. To this end we solve the evolution
equation \eqref{s-space-evol} numerically, using in one case two-loop $\Gamma_{\cusp}$ and one-loop
$\gamma_\eta$, and in another case three-loop $\Gamma_{\cusp}$, two-loop
$\gamma_\eta$ and the mixing term $\mathcal{O}(a^2)$ on the r.h.s. of \eqref{s-space-evol}.
We refer to these truncations as the next-to-leading-logarithmic (NLL) and the
next-to-next-to-leading logarithmic (NNLL) resummation, respectively. In both cases we use
three-loop QCD coupling.

We see that the NNLL correction is in general small, which is consistent
with the observation in Ref.~\cite{Beneke:2018wjp} that dependence
of the $B\to\ell\nu_\ell\gamma$ form factors on the hard-collinear factorization scale is rather weak.
The correction is negative at small momenta, and positive at large momenta.
This is also true for more general models considered in~\cite{Beneke:2018wjp} although the
size of the correction at small momenta can be larger if the lower-energy LCDA does not
have the linear behavior at $\omega\to 0$ expected in perturbation theory.

For the leading-power contribution in QCD factorization, the precise
functional form of the LCDA is not important as the result can be expressed in
terms of the logarithmic moments~\cite{Beneke:2018wjp}
\begin{align}
\widehat{\sigma}_n = \int^\infty_0 d\omega\,\frac{\lambda_B}{\omega}\,
\ln^n\frac{\lambda_B e^{-\gamma_E}}\omega\,\phi_+(\omega)
\label{sigma-n}
\end{align}
with $\widehat \sigma_0 =1$ defining $\lambda_B$. To the NNLL accuracy only the
values of $\lambda_B$, $\hat \sigma_1$ and $\hat \sigma_2$ are needed.
For the simple model in Eq.\,\eqref{GNmodel} $\lambda_B(\mu_0) = \omega_0$, $\widehat{\sigma}_1(\mu_0)=0$.
After the evolution to 2 GeV one obtains, for three typical parameter values:
\begin{center}
\begin{tabular}{|c@{\hspace{0.5cm}}|c@{\hspace{0.5cm}}|c@{\hspace{0.5cm}}|c@{\hspace{0.5cm}}|c@{\hspace{0.5cm}}|}
\hline
 $\omega_0,\text{MeV}$ & $\lambda_B^{\scriptscriptstyle \rm NLL}/\omega_0 $ & $\lambda_B^{\scriptscriptstyle \rm NNLL}/\omega_0$ &
                         $\widehat{\sigma}_1^{\scriptscriptstyle \rm NLL}$ & $\widehat{\sigma}_1^{\scriptscriptstyle \rm NNLL}$\\
\hline
  200       &  1.29                 &  1.31           & \phantom{-}0.011        &    -0.042   \\
  300       &  1.22                 &  1.24           & -0.043                  &    -0.116   \\
  400       &  1.18                 &  1.18           & -0.082                  &    -0.172   \\
\hline
\end{tabular}
\end{center}
More detailed numerical studies should be done in
connection with concrete physics applications.

To summarize, we have studied higher-order corrections to the scale-dependence of the
B-meson LCDA. We reveal the general structure of the evolution kernel and its
relation to conformal symmetry of QCD Lagrangian, and confirm this structure by explicit
two-loop calculation. The resulting evolution equation \eqref{s-space-evol} is the
last missing ingredient that allows one to to perform QCD factorization in
charmless B decays to the NNLL accuracy.\\


{\large\bf Acknowledgments:}~~~
We thank Jakob Sch{\"o}nleber for participation in a part of relevant calculations and Alexey Vladimirov
for helpful comments on the manuscript.
This work was supported by the DFG grants BR 2021/7-2 (YJ), MO 1801/1-3 (AM), and the RSF project 19-11-00131 (AM).



\bibliography{ref-NLO}



\end{document}